\documentclass{emulateapj} \usepackage{psfig} \usepackage{apjfonts}

\newcommand\beq{\begin{equation}}
\newcommand\eeq{\end{equation}}
\def\rmmat#1{{\hbox{\rm #1}}}

\newcommand{\etal}{et al.}

\newcommand\xmm{{\it XMM-Newton}}

\def\taxp{\hbox{XTE~J1810$-$197}}

\begin{document}
 
\title{Constraints on the Emission and Viewing Geometry of \\ the Transient Anomalous X-ray Pulsar XTE J1810$-$197}

\author{Rosalba Perna\altaffilmark{1} and E.~V.~Gotthelf\altaffilmark{2}} 
\altaffiltext{1}{JILA and Department of Astrophysical and Planetary Sciences, 
University of Colorado, Boulder, CO, 80309}
\altaffiltext{2}{Columbia Astrophysics Laboratory, Columbia University, New York, NY 10027}

\begin{abstract}
The temporal decay of the flux components of Transient Anomalous X-ray
Pulsar \taxp\ following its 2002 outburst presents a unique
opportunity to probe the emission geometry of a magnetar. Toward this
goal, we model the magnitude of the pulsar's modulation in narrow
spectral bands over time.  Following previous work, we assume that the
post-outburst flux is produced in two distinct thermal components arising
from a hot spot and a warm concentric ring. We include general
relativistic effects on the blackbody spectra due to gravitational
redshift and light bending near the stellar surface, which strongly
depend on radius. This affects the model fits for the temperature and
size of the emission regions. For the hot spot, the observed temporal
and energy-dependent pulse modulation is found to require an
anisotropic, {pencil-beamed} radiation pattern. We are able to
constrain an allowed range for the angles that the line-of-sight
($\psi$) and the hot spot pole ($\xi$) make with respect to the
spin-axis. Within errors, this is defined by the locus of points in
the $\xi-\psi$-plane that lie along the line
$(\xi+\beta(R))(\psi+\beta(R))\approx{\rm constant}$, where $\beta(R)$
is a function of the radius $R$ of the star.  For a canonical value of
$R=12$~km, the viewing parameters range from $\psi=\xi=37^{\circ}$ to
$(\psi,\xi)=(85^{\circ},15^{\circ})$. We discuss our results in the
context of magnetar emission models.

\end{abstract}

\keywords{pulsars: individual (\taxp) --- stars: neutron --- X-rays: stars}

\section{Introduction}

Anomalous X-ray Pulsars (AXPs) are peculiar high-energy pulsars whose
observed luminosity greatly exceeds that which can be supplied by
their rotational energy losses. These pulsars occupy a narrow range of
spin periods ($P\sim 2-12$ sec) and are spinning down rapidly compared
to the rotation-powered pulsars. For the vacuum dipole model, the
timing properties of these pulsars imply an enormous magnetic field,
$B\sim 10^{14}-10^{15}$ G.  These relatively rare objects ($\sim 10$
compared to $\sim 1700$ catalogued radio pulsars), generally display
sinusoidal modulation in their pulsed flux, with a wide range of
amplitudes ($\sim 10-80\%$) and are likely young ($<10^4$~yrs), as more
than half are associated with supernova remnants (see Kaspi 2007 for a
recent review). AXPs can be understood within the context of the
magnetar model developed by Duncan \& Thompson (1992) to explain the
burst phenomenology of Soft $\gamma$-ray Repeaters (SGRs). The excess
emission from both AXPs and SGRs, collectively referred to as
magnetars, is powered by the decay of their extreme magnetic
fields. This is suggested by the relatively high temperatures of their
thermal emission ($kT \approx 0.4-0.7$~keV for blackbody fits), and
frequent rapid ($<0.1$~s) burst activity. The geometry and the
properties of the observed emission from the magnetars is of great
interest for understanding how this activity arises.

The recent discovery of an AXP fading from a long duration outburst
offers the unique opportunity to probe the magnetar emission geometry
evolution during this event. The $P=5.54$~s Transient AXP (TAXP)
\taxp\ was discovered in January 2003 by Ibrahim et al. (2004) using
the {\em Rossi X-ray timing explorer} ({\it RXTE}) following a large
eruption. Subsequently its flux decayed exponentially ($\tau \approx
900$~d) nearly back to a quiescent flux level as determined from
serendipitously archival X-ray observations (Gotthelf et
al. 2004). The earlier measurements indicate that the TAXPs $2-10$~keV
flux had increased by two orders of magnitude. However, of great
interest, the quiescent luminosity is 100 times lower than for the
well-established AXPs and SGRs, suggesting a large unidentified
population of neutron stars (Gotthelf \etal\ 2004). In contrast, the
magnetic field strength of \taxp, $B = 3\times 10^{14}$~G as inferred
from its spin-down properties, is typical of the magnetars.

The flux and pulse evolution of \taxp\ were monitored with the \xmm\
X-ray observatory at roughly bi-yearly intervals starting Sept. 2003,
yielding a total of seven epochs through Mar. 2006. The complete set of
observations, together with their spectral modeling and
interpretation, is described in detail by Gotthelf \& Halpern (2007),
with the earlier observations reported in Gotthelf \& Halpern (2005)
and Halpern \& Gotthelf (2005). While analysis of phase-averaged
spectra alone cannot distinguish among competing models for the AXP
emission type and geometry, the addition of the steady change of the
spectrum and pulse modulation over time greatly increases the
diagnostic power.

In this paper we present a detailed model for the energy dependent
pulse phase from \taxp.  This model accounts for the viewing
geometry and surface emission distribution. We include the general
relativistic effects of light deflection and gravitational redshift
and allow for anisotropic emission. We apply this model to a set of
X-ray data acquired during the temporal evolution of the flux from
\taxp. This allows us to constrain the underlying emission
geometry and radiation properties of this transient magnetar.

\section{Time-dependent Flux Modeling of \taxp}

\subsection{Model Motivations}

Since their discovery, spectra of magnetars have been fitted with a
variety of models, generally including two components, such as
blackbody plus power-law, atmosphere plus power-law (Perna et
al. 2001a; Skinner et al. 2006), or thermal plus resonant cyclotron scattering (Rea et
al. 2007), or more recently, a magnetized atmosphere model with the
inclusion of scattering (Guver et al. 2007).  From the point of view
of phase-averaged spectral analysis alone, these models are generally
statistically acceptable, and therefore none can be ruled out {\it a
priori}.

Phase-resolved modeling of the observed modulation, however, can
provide a much stronger constraint. This is particularly true for the
case of \taxp\ , given the wealth of data available at different
epochs, while the object is cooling. From the point of view of being
able to reproduce the observed time-dependent energy behavior of the
pulsed fractions, the double blackbody model (made up of a hot and a
warm component) put forward by Gotthelf \& Halpern (2005) is quite
promising, at least from a qualitative point of view.  In fact, if, in
the same energy band, the warm component is less pulsed than the hot
one, then, as the hot component drops in flux faster than the warm one
(as found in their fitting), the pulse fraction will tend to decline,
as observed.  Also, in their modeling, the area of the warm component,
which becomes more luminous than the hot component after the fourth
epoch, increases at later times.\footnote{This area increase might
either be real, or a result of the fact that the warm component cannot
be straightforwardly separated in the fits from the underlying surface
emission, to which it approaches at later times. Either way, the pulse
modulation in this model is expected to decrease with time, since the
surface emission of the neutron star, if its temperature distribution
traces that of a dipolar magnetic field, is not expected to be highly
modulated (DeDeo et al. 2001).}  This, again, tends to produce a
decline in the pulse modulation and counteract the increase that would
otherwise have, due to the decreasing temperature.

While it is tempting to model the thermal components with detailed
magnetized atmospheres (e.g. van Adelsberg \& Lai 2006), these models
might be problematic for the case of \taxp\ following its
outburst. Invoking dissipation of a twist in the magnetic field lines
(Beloborodov \& Thompson 2007), the field in the emitting region is
likely to have significant non-normal components.  While the ``twist
model'' nicely predicts the timescale of the outburst decay, the
non-{ normal surface} magnetic fields ({ i.e. magnetic fields that
emerge from the star surface at oblique angles}) have yet to be fully
realized in the magnetized atmosphere models.  Since the predicted
amplitude of the flux modulation  strongly depends on the local
magnetic field direction, by assuming a magnetized atmosphere model
for the thermal components we would introduce an {\it a priori} bias
in our results.

In this work, we prefer to take a more empirical approach by starting
with the distribution of the emitted radiation over the stellar
surface and allowing a degree of anisotropy (beaming factor) in the
thermal (blackbody) components (following the methods of Pechenick et
al. 1993; DeDeo et al. 2000; Perna et al. 2001b), and leaving the
beaming factor as one of the model parameters. This approach turns out
to be very valuable in that we can use the energy-dependency of
the pulsed fractions, together with their variation with time as the
emitting region cools, to constrain both the viewing geometry and the
beaming properties of the radiation simultaneously (\S 3). Our results
can therefore be used as a guide for further theoretical modeling
aimed at understanding the detailed mechanisms that produce the
observed thermal radiation.

\subsection{The Emission Model}

Our method for parameterizing the surface emission from \taxp\
follows the example developed by Pechenick et al. (1993) with some
generalizations.  For our numerical modeling we consider emission from
a hot spot of temperature $T_h$ and angular radius $\beta_h$
surrounded by a warm ring of temperature $T_w$ and outer radius $\beta_w$

%
%

\begin{figure}[t]
\plotone{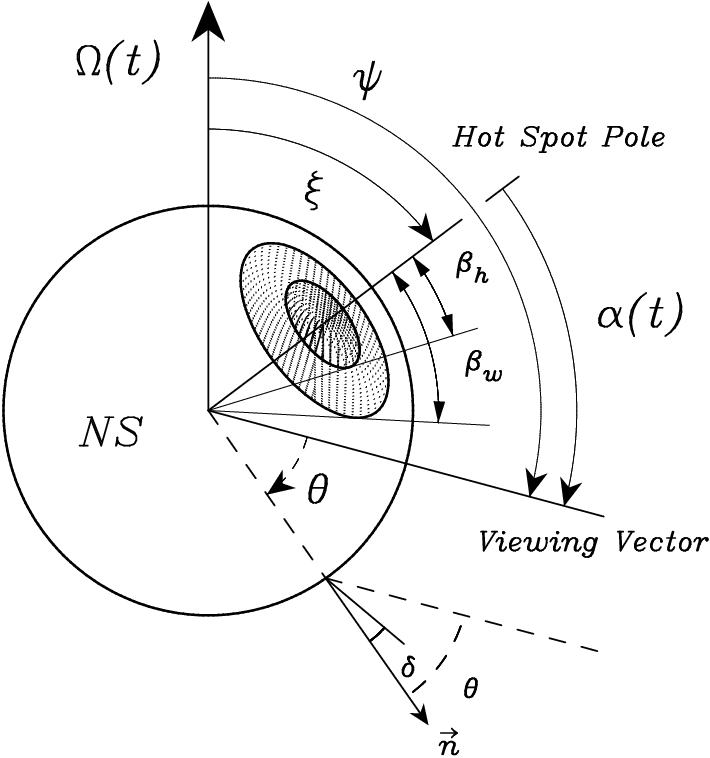}
\caption{Emission geometry on the surface of the neutron star (NS) for
the model presented herein. A hot spot of temperature $T_h$ and
angular size $\beta_h$ is surrounded by a warm ring of temperature
$T_w$ with outer angular size $\beta_w$. As the neutron star (NS)
rotates with angular velocity $\Omega(t)$, the angle $\alpha(t)$ is a
function of the phase angle $\gamma(t)=\Omega(t)t$, and the angles
$\psi$ and $\xi$ between spin axis and viewing vector, and between
spin axis and emission pole, respectively. { Due to general relativistic effects,
a photon emitted at a colatitude $\theta$ on the star's surface
which reaches the observer must be emitted at an angle $\delta$ with respect to the star's normal
at that point.} }
\vspace{0.1in}
\label{fig:NS}
\end{figure}

We indicate with $\alpha$ the angle that the axis of the hot spot
makes with respect to the line of sight. This depends on the phase
angle $\gamma(t)=\Omega(t) t$, as the star rotates with
angular velocity $\Omega(t)$. If $\xi$ is the angle between the spot
axis and the rotation axis, and $\psi$ the angle between the
observer's direction and the rotation axis, then the angle $\alpha$ is
given by 

\beq
\alpha(t)=\arccos(\cos\psi\cos\xi+\sin\psi\sin\xi\cos\gamma(t))\;.
\label{eq:alpha}
\eeq

The geometry is illustrated in Figure 1. The surface of the star is
described by the angular spherical coordinates $(\theta,\phi)$, and
the coordinate system is chosen so that the $z$ axis is in the
direction of the line of sight to the observer. {This is a natural
choice of coordinate system for our problem, since the observed flux
is produced by all the photons that reach the observer at $\infty$
along that axis (e.g.  Pechenick et al. 1993; Page 1995). Note that,
if there were no general relativistic effects, a photon emitted at
colatitude $\theta$ on the star would only reach the observer if it
were emitted at an angle $\delta=\theta$ with the normal to the
surface of the star. Because of general relativistic effects, however,
a photon emitted at a colatitude $\theta$ will get to the observer
only if emitted at an angle $\delta$ with respect to the surface normal (see
Fig.1), where the relation between the two angles is given by the
ray-tracing function\footnote{To improve the computational efficiency
of the above equation we use the approximation presented in
Beloborodov (2002).} (Pechenick et al. 1993; Page 1995)} \beq
\theta(\delta)=\int_0^{R_s/2R}x\;du\left/\sqrt{\left(1-\frac{R_s}{R}\right)
\left(\frac{R_s}{2R}\right)^2-(1-2u)u^2 x^2}\right.\;,
\label{eq:teta}
\eeq 
\noindent having defined $x\equiv\sin\delta$. Here, $R/R_s$ is the ratio of the
stellar to Schwarzschild radius, $R_s=2GM/c^2$ (we assume $M=1.4
M_\odot$). 

The hot spot is described by
the conditions: 

\beq
\theta\le\beta_h,\;\;\;\;\;\;\;\;\;\;\;\; \rmmat{if}\;\;\; \alpha=0\;
\label{eq:con1}
\eeq
and
\beq
   \left\{
  \begin{array}{ll}
    \alpha-\beta_h\le\theta\le\alpha+\beta_h \\
      2\pi-\phi_p^h\le\phi\le\phi_p^h \;\; \;\;\;\;\rmmat{if}
        \;\;\;\alpha\ne 0\;\;\;\rmmat{and} \;\;\;\beta_h\le\alpha\\
  \end{array}\right.\;
\label{eq:con2}
\eeq
where
\beq
\phi_p^h=\arccos\left[\frac{\cos\beta_h-\cos\alpha\cos\theta}{\sin\alpha\sin\theta}\right]\;.
\label{eq:phip}
\eeq

On the other hand, it is identified through the condition 

\beq
\theta\le\theta^h_*(\alpha,\beta_h,\phi),\;\;\;\;\ 
\rmmat{if}\;\;\; \alpha\ne 0\;\;\;\rmmat{and}\;\;\;\beta_h > \alpha\;,
\label{eq:con3}
\eeq

\noindent where the outer boundary $\theta^h_*(\alpha,\beta_h,\phi)$ of the spot is computed by numerical
solution of the equation 

\beq
\cos\beta_h = \sin\theta_*^h\sin\alpha\cos\phi + \cos\theta_*^h\cos\alpha\;.
\label{eq:t*}
\eeq

Similarly, the warm ring is described on the star surface through the
conditions

\beq
\beta_h < \theta\le\beta_w,\;\;\;\;\ \rmmat{if}\;\;\; \alpha=0\;
\label{eq:con1w}
\eeq

and

\beq
   \left\{
  \begin{array}{ll}
    \alpha-\beta_w\le\theta\le\alpha+\beta_w \;\;\;\rmmat{and} 
\;\;\;\alpha+\beta_h\le\theta\le\alpha-\beta_h\\
      2\pi-\phi_p^w\le\phi\le\phi_p^w \;\;\;\rmmat{and}
\;\;\;\phi_p^h\le\phi\le 2\pi - \phi_p^h \\
  \end{array}\right.\;
\label{eq:con4}
\eeq 

\noindent if $\alpha\ne 0$ and $\beta_w\le\alpha$. In the above equation,
$\phi_p^w$ has the same functional form as $\phi_p^h$ in
Eq~(\ref{eq:phip}), except for the substitution
$\beta_h\rightarrow\beta_w$.  Finally, if $\alpha\ne 0$ and $\beta_w >
\alpha$, the ring is identified by the condition 

\beq
\theta^h_*(\alpha,\beta_h,\phi)<
\theta\le\theta^w_*(\alpha,\beta_w,\phi), 
\label{eq:t*w}
\eeq 

\noindent where, again, the outer boundary of the ring
$\theta^w_*(\alpha,\beta_w,\phi)$ is found by numerical solution of
Eq.~(\ref{eq:t*}), but with the replacement
$\beta_h\rightarrow\beta_w$.

In the following we assume a blackbody emission model for both the hot
spot and warm ring, characterized by a uniform temperature ($T_h$ or
$T_w$, respectively) over their stellar surface.  As discussed above,
we allow for the radiation from the two regions to be anisotropic, and
parameterize the beaming of their {local} emission through the
functions $f_{h}(\delta)\propto \cos^{n_h}(\delta)$ (hot spot) and
$f_{w}(\delta)\propto \cos^{n_w}(\delta)$ (warm ring). { This
choice was initially motivated by the consideration that the hot spots
are likely associated with regions of larger conductivity, where the
magnetic field lines would be close to perpendicular to the surface of the
star. This would produce an enhanced emissivity at small $\delta$. Our
analysis (\S3.2) then confirmed the validity of this choice by
demonstrating that the modulation of the hottest region does indeed
require a pencil-type anisotropic beaming pattern.}

The observed spectrum as a function of phase angle $\gamma$ is then
obtained by integrating the local emission over the observable surface
of the star, accounting for the gravitational redshift of the
radiation (Page 1995)

\begin{eqnarray}
F(E_\infty,\gamma) =\frac{2 \pi}{c\,h^3}\frac{R_\infty^2}{D^2}\;E_\infty^2
e^{-N_{\rm H}\sigma(E_\infty)} \int_0^1 2xdx\nonumber \\ 
 \times \int_0^{2\pi} \frac{d\phi}{2\pi}\; 
I_0(\theta,\phi) \;n[E_\infty e^{-\Lambda_s};T(\theta,\phi)]\;,
\label{eq:flux}
\end{eqnarray} 

\noindent in units of photons cm$^{-2}$ s$^{-1}$ keV$^{-1}$.  In the above
equation, the radius and energy as observed at infinity are given by
$R_\infty= Re^{-\Lambda_s}$, and $E_{\infty}= E e^{\Lambda_s}$, where
$R$ is the star radius, $E$ is the energy emitted at the star surface,
and we have defined ${\Lambda_s}$ as, 

\beq
e^{\Lambda_s}\equiv\sqrt{1-{\frac{R_s}{R}}}.  
\eeq 

For the spectral function, given by $n(E,T)=1/[\exp(E/kT)-1]$, the
temperature $T(\theta,\phi)$ is equal to $T_h$ if \{$\theta, \phi$\}
satisfy any of the conditions (\ref{eq:con1}) through (\ref{eq:t*}),
and it is given by $T(\theta,\phi)= T_w$ if any of the conditions
(\ref{eq:con1w}) through (\ref{eq:t*w}) holds true.  Correspondingly,
the weighted intensity $I_0(\theta,\phi)$ is given by the functions
$f_h$ or $f_w$ depending on whether the variables \{$\theta,\phi$\}
are inside the hot or warm region, respectively.

The phase-averaged flux is readily computed as $F_{\rm ave}(E_\infty)=
1/2\pi\int_0^{2\pi}d\gamma F(E_\infty,\gamma)$. The phase dependence
$\gamma$ in Eq.(\ref{eq:flux}) comes from the viewing angles implicit
in $\alpha(t)$ and from the series of conditions (\ref{eq:con1})
through (\ref{eq:t*w}).  {Note that, as the star rotates, the only
angle on which the flux depends is $\alpha(t)$, the angle that the
line of sight makes with the axis of the spots. Since, in the magnetar
model, the spots are likely to be correlated to regions with an
enhanced magnetic activity, the angle $\alpha(t)$ can also be
considered as the (phase-dependent) angle between the line of sight
and a magnetically active region on the star during and following the
outburst. When the star returns to quiescence, the temperature
distribution on the star will reflect the overall magnetic field
configuration. For most AXPs, the quiescent emission cannot be
produced by a temperature distribution following a dipolar magnetic
field (De Deo, Psaltis \& Narayan 2000); in the case of XTE~J1810-197,
a detailed study of the quiescent emission, once the contamination
from the heated regions has completely subsided, will be able to
determine the detailed structure of the surface temperature
distribution, and hence reconstruct the magnetic field structure in
quiescence.  This study will be performed in a forthcoming
paper. However, a preliminary investigation of the softest energy band
in the latest data set (where the surface emission from the rest of
the star is likely to be dominant) shows that the maximum of the
pulsed emission remains in phase with the maximum in the hardest
energy band (still dominated by the heated region). This result shows
that the maximum of the quiescent emission comes from the region where
the outburst occurred. Therefore, the active region is likely to be
associated with an enhanced magnetic field strength in quiescence. If
the underlying $B$ field is dipolar (or close to such), then the spot
axis in our paper also represents the dipole magnetic field axis, and
therefore the angle $\alpha(t)$ with respect to the $z$-axis
(observer's viewing direction) would naturally be associated with the angle that
the magnetic axis forms with respect to our line of sight as the star rotates.}

\begin{deluxetable}{lccccccc}
\tablecolumns{8}
\tighten
\tablewidth{0.0pt}
\tablecaption{\bf Spectral Results as a Function of the NS Radius - Model Fits for $\bf N_{\rm{H}}=6.8 \times 10^{21}$~cm$^{-2}$, $\bf D=3.3$~kpc, ${\bf \psi=\xi=\gamma=0^{\circ}}$ (see text for definitions)\label{tab:spectable}}
\tablehead{
\colhead{Parameter} &  \multicolumn{2}{c}{2003}  & \multicolumn{2}{c}{2004} & \multicolumn{2}{c}{2005} & \colhead{2006}\\ 
\colhead{}          &  \colhead{Sep 8} &  \colhead{Oct 12}  & \colhead{Mar 11} & \colhead{Sep 18} & \colhead{Mar 18} & \colhead{Sep 20} & \colhead{Mar 12} 
}
\startdata
\cutinhead{$R=9$~km; $\chi^2_{\nu}(\rm{dof}) = 1.09(1914)$ } 
$kT_h$ (keV)	      & $0.91\phantom{}$      & $0.95\phantom{}$      & $0.93\phantom{}$     & $0.91\phantom{}$     & $0.84\phantom{}$      & $0.71\phantom{}$      & $0.60\phantom{}$\\
$kT_w$ (keV)          & $0.32\phantom{}$      & $0.38\phantom{}$      & $0.35\phantom{}$     & $0.35\phantom{}$     & $0.31\phantom{}$      & $0.26\phantom{}$      & $0.23\phantom{}$\\
$\beta_h$ (deg)	      & $10.06 \phantom{}$   & $8.47\phantom{}$   & $6.69\phantom{}$  & $5.33\phantom{}$  & $3.81\phantom{}$   & $3.52\phantom{}$   & $4.08\phantom{}$ \\
$\beta_w$ (deg)       & $56.3\phantom{}$   & $38.3\phantom{}$   & $39.3\phantom{}$  & $42.3\phantom{}$  & $42.3\phantom{}$   & $55.1\phantom{}$   & $82.1\phantom{}$ \\
\cutinhead{$R=10$~km;  $\chi^2_{\nu}(\rm{dof}) = 1.11(1914)$}
$kT_h$ (keV)	      & $0.91\phantom{}$      & $0.95\phantom{}$      & $0.890\phantom{}$     & $0.91\phantom{}$     & $0.76\phantom{}$      & $0.67\phantom{}$      & $0.59\phantom{}$\\
$kT_w$ (keV)          & $0.36\phantom{}$      & $0.40\phantom{}$      & $0.33\phantom{}$     & $0.35\phantom{}$     & $0.27\phantom{}$      & $0.24\phantom{}$      & $0.22\phantom{}$\\
$\beta_h$ (deg)	      & $8.21 \phantom{}$   & $7.21\phantom{}$   & $5.83\phantom{}$  & $4.46\phantom{}$  & $4.03\phantom{}$   & $3.12\phantom{}$   & $3.11\phantom{}$ \\
$\beta_w$ (deg)       & $35.8\phantom{}$   & $29.4\phantom{}$   & $34.6\phantom{}$  & $30.2\phantom{}$  & $43.8\phantom{}$   & $51.9\phantom{}$   & $66.5\phantom{}$ \\
\cutinhead{$R=11$~km;  $\chi^2_{\nu} (\rm{dof}) = 1.15(1914)$}
$kT_h$ (keV)	      & $0.89\phantom{}$      & $0.90\phantom{}$      & $0.87\phantom{}$     & $0.83\phantom{}$     & $0.75\phantom{}$      & $0.67\phantom{}$      & $0.56\phantom{}$\\
$kT_w$ (keV)          & $0.35\phantom{}$      & $0.37\phantom{}$      & $0.32\phantom{}$     & $0.29\phantom{}$     & $0.27\phantom{}$      & $0.24\phantom{}$      & $0.21\phantom{}$\\
$\beta_h$ (deg)	      & $7.05 \phantom{}$   & $6.98\phantom{}$   & $5.25\phantom{}$  & $4.37\phantom{}$  & $3.44\phantom{}$   & $2.85\phantom{}$   & $3.27\phantom{}$ \\
$\beta_w$ (deg)       & $31.3\phantom{}$   & $27.8\phantom{}$   & $31.5\phantom{}$  & $36.3\phantom{}$  & $37.1\phantom{}$   & $43.8\phantom{}$   & $61.3\phantom{}$ \\
\cutinhead{$R=12$~km;  $\chi^2_{\nu} (\rm{dof}) = 1.09(1914)$}
$kT_h$ (keV)	      & $0.85\phantom{}$  & $0.86\phantom{}$      & $0.86\phantom{}$     & $0.82\phantom{}$     & $0.71\phantom{}$      & $0.63\phantom{}$      & $0.59\phantom{}$\\
$kT_w$ (keV)          & $0.32\phantom{}$      & $0.34\phantom{}$      & $0.32\phantom{}$     & $0.29\phantom{}$     & $0.25\phantom{}$      & $0.24\phantom{}$      & $0.22\phantom{}$\\
$\beta_h$ (deg)	      & $7.06 \phantom{}$   & $6.39\phantom{}$   & $5.01\phantom{}$  & $4.22\phantom{}$  & $3.45\phantom{}$   & $2.53\phantom{}$   & $2.27\phantom{}$\\
$\beta_w$ (deg)       & $33.8\phantom{}$   & $28.1\phantom{}$   & $28.6\phantom{}$  & $32.1\phantom{}$  & $37.8\phantom{}$   & $41.2\phantom{}$   & $45.3\phantom{}$\\
\cutinhead{$R=13$~km;  $\chi^2_{\nu} (\rm{dof}) = 1.10(1914)$}
$kT_h$ (keV)	      & $0.83\phantom{}$      & $0.85\phantom{}$      & $0.82\phantom{}$     & $0.76\phantom{}$     & $0.69\phantom{}$      & $0.61\phantom{}$      & $0.53\phantom{}$\\
$kT_w$ (keV)          & $0.31\phantom{}$      & $0.33\phantom{}$      & $0.29\phantom{}$     & $0.26\phantom{}$     & $0.24\phantom{}$      & $0.23\phantom{}$      & $0.20\phantom{}$\\
$\beta_h$ (deg)	      & $6.35 \phantom{}$   & $5.84\phantom{}$   & $4.81\phantom{}$  & $4.14\phantom{}$  & $3.29\phantom{}$   & $2.44\phantom{}$   & $2.83\phantom{}$ \\
$\beta_w$ (deg)       & $31.4\phantom{}$   & $26.5\phantom{}$   & $29.6\phantom{}$  & $36.7\phantom{}$  & $37.9\phantom{}$   & $39.2\phantom{}$   & $49.2\phantom{}$ \\
\cutinhead{$R=14$~km;  $\chi^2_{\nu} (\rm{dof}) = 1.14(1914)$}
$kT_h$ (keV)	      & $0.80\phantom{}$      & $0.81\phantom{}$      & $0.79\phantom{}$     & $0.74\phantom{}$     & $0.67\phantom{}$      & $0.59\phantom{}$      & $0.52\phantom{}$\\
$kT_w$ (keV)          & $0.28\phantom{}$      & $0.30\phantom{}$      & $0.27\phantom{}$     & $0.23\phantom{}$     & $0.23\phantom{}$      & $0.22\phantom{}$      & $0.19\phantom{}$\\
$\beta_h$ (deg)	      & $6.16 \phantom{}$   & $5.79\phantom{}$   & $4.49\phantom{}$  & $4.07\phantom{}$  & $3.23\phantom{}$   & $2.40\phantom{}$   & $2.45\phantom{}$ \\
$\beta_w$ (deg)       & $33.26\phantom{}$   & $28.9\phantom{}$   & $30.7\phantom{}$  & $43.5\phantom{}$  & $38.5\phantom{}$   & $38.1\phantom{}$   & $47.9\phantom{}$\\
\enddata
\tablecomments{Uncertainties on each spectral parameter are consistent with those reported for the fits in Table 3.}
\end{deluxetable}

\section{Modeling the Multi-epoch Modulation}

The above model was coded in {\tt FORTRAN} and fully implemented as an
additive model with 10 parameters in the {\tt XSPEC} spectral fitting
software (Arnaud 1996). Specifically, these parameters are the
temperatures ($kT_h, kT_w$; keV) and subtended angular sizes
($\beta_h, \beta_w$; degs) of the hot spot and warm ring,
respectively, the viewing ($\psi$) and hot spot ($\xi$) angles w.r.t
the spin-axis (in degrees), the rotation phase ($\gamma$; cycles), and
finally, the NS radius ($R$; km) and distance ($D$; kpc).  The {\tt
XSPEC} normalization is set to unity so that the flux is fixed by the
distance and stellar radius, which implicitly takes into account all
relativistic effects previously noted in \S2.  In the following
spectral fits the pulsar distance is set to $D=3.3$~kpc, based on
radio pulse dispersion (Camilo et al. 2006), and consistent with the
measurement derived from HI absorption (Minter et al. 2007).

This model allows us to predict the energy dependent modulation, and
use it to determine the viewing geometry and beaming pattern of the
emitted radiation that best match the observations at different
epochs. We assume both to be temporal invariant, i.e. no noticeable
precession changes with time.  In principle, the neutron star radius,
because of general relativistic effects, could be uniquely
determined; however, in practice the presence of noise does not allow
for this most interesting of constraints.  In the following, we apply
this model to the data sets presented in Gotthelf \& Halpern (2007),
which fully described their preparation. We first present the spectral
fits (\S3.1) using the full model, assuming a face-on geometry
($\alpha =0$), and then constrain the overall emission geometry of the
system (\S3.2) by modeling the observed pulse modulation in 6 energy
bands over time.

All spectral fitting are done in the $0.7-10$~keV spectral band
assuming no beaming initially, as this is not an important effect
spectrally. However, some degree of anisotropy of the radiation is
found to be necessary in our model to reproduce the observed
modulation (\S3.2).

\subsection{Spectral analysis} 

We started by fitting the phase-averaged \xmm\ spectra for the 7
epochs simultaneously using our model for the pulsar emission
geometry. Since the viewing geometry is not known {\it a priori}, we
assume the simplest choice, that we are looking directly down the
co-aligned rotation axis and magnetic pole ($\alpha=0$, see
Eq.~1). This has the practical benefit of allowing the model code to
run substantially faster since, for this special case of $\alpha=0$,
the integration is simpler and only one call to the routine is needed
for the computation of the phase-averaged spectrum. Across all epochs,
all parameters are linked with the exception of the set of 4 epoch
variable parameters ($kT_w, kT_h, \beta_w, \beta_h$).  Initial fits
were used to determine the nominal column density of $N_{\rm
H}=6.8\times 10^{21}$ cm$^{-2}$, which was subsequently fixed to this
value.

An important technical issue for these fits is the degree of
degeneracy between the radius $R$ and the four epoch variable
parameters ($kT_w, kT_h, \beta_w, \beta_h$). These 5 parameters
over-determine the fit, unlike fits using a double blackbody model.
Without fixing the radius there is no unique solution, and thus we
consider a range of possible values between $9\le R \le 14$~km, in
1~km increments. These results are presented in Table~1 and show a
similar trend to those reported by Gotthelf \& Halpern (2007) using
the double-blackbody model. In both cases, the hot components is found
to steadily decrease in size over time, while the warm component
increases (with the exception of the first data point).

In our model, the radius of the star is not just a simple
normalization. This is due to the introduction of gravitational
redshift effects. Unlike for the non-relativistic case, the inferred
temperatures of the spots increase as the radius of the star becomes
smaller. Two counteracting effects, both due to flux conservation
influence the spot size -- gravitational redshift tends to decrease
the inferred emission area in the more relativistic (smaller) stars
(due to the higher inferred temperatures); on the other side, for a
fixed distance { between the star and the observer}, the spot
angular size increases on smaller stars. For the values of the fit
parameters here, the latter effect tends to dominate over the
former. Over the sampled range, we do not find evidence for a
preferred radius, based on the $\chi^2$ measurements.

As discussed above, the results of our spectral fits using the above model show a similar
trend to those reported by Gotthelf \& Halpern (2007) using a
double-blackbody model.  The hot components steadily decrease in size
over time, while the warm component increases (except for the first
data point).  We are aware of the importance of a possible third
emission component from the rest of the NS surface, perhaps the
quiescent emission, initially masked by the significant extra flux
from the warm component activated by the outburst.  However, we are
unable to resolve any additional component, which is not required by
the spectral fits. If the last few data sets are substantially
affected by this potential third component, our pulse profile modeling
of those data could be incomplete.  Therefore, for the second part of
our analysis, we rely on the first 4 data sets alone, during which the
emission from the two components dominates over that from the NS
surface (whose quiescent level was measured with ROSAT prior to the
outburst).

\begin{deluxetable}{ccccccc}
\tablecolumns{7}
\tighten
\tablewidth{0.0pt}
\tablecaption{\bf  Minimum $\chi^2_\nu$ as a Function of Beam Indices ({$\bf n_w,n_h$}) and NS Radius\label{tab:beamtable}}
\tablehead{
\colhead{$n_w,n_h$} &  \multicolumn{6}{c}{Minimum  $\chi^2_{\nu}$} \\
                    & \colhead{$R=$9~km} & \colhead{$R=$10~km} & \colhead{$R=$11~km} & \colhead{$R=$12~km} & \colhead{$R=$13~km} & \colhead{$R=$14~km}
}
\startdata
$0,0$ & 4.06 & 4.91 & 3.97 & 4.62 & 3.85 & 2.86\\
$0,1$ & 0.86 & 1.07 & 1.13 & 0.96 & 0.94 & 0.99\\
$0,2$ & 3.17 & 3.25 & 3.74 & 3.02 & 3.29 & 2.83\\
$1,1$ & 2.51 & 3.50 & 2.55 & 2.64 & 1.81 & 1.17
\enddata
\tablecomments{Minimum reduced
$\chi^2_{\nu}$ after comparing model and observed PFs over the
$\xi,\psi$-space, with the NS radius and beaming indices held fixed at
the given values. Only the first 4 epochs were included in this analysis.}
\end{deluxetable}

\subsection{Modulation analysis}

Starting with the best spectral fit model parameters presented in
Table~1, obtained for $\xi=\psi=0$, we now searched for the best
values of $\xi$ and $\psi$ needed to reproduce the observed magnitude
of the pulse modulation across the first 4 epochs, as measured by
Gotthelf \& Halpern (2007). Given the uncertainties in the data, we
limit our modeling to that of the pulse modulation, rather than the
full pulse profile. The pulsed fractions were determined in six energy
bands\footnote{\{0.5-1; 1-1.5; 1.5-2; 2-3; 3-5; 5-8\} keV.} at each
epoch.  For each value of the NS radius ($R$) fitted for in \S3.1, we
computed the modulation (defined below) over the grid of angles
$\xi,\phi\le90$ deg, in 1 degree intervals.  For each value on the
grid, the model predictions were compared with the data.  Notice that
the flux depends on the angles $\xi$ and $\psi$ only through the
parameter $\alpha$ in Eq.(\ref{eq:alpha}), and therefore it is
symmetric with respect to an exchange of $\xi$ and $\psi$. The
magnitude of the model modulation is defined as

\begin{equation}
PF=\frac{F_{\rm max}-F_{\rm min}}{F_{\rm max}+F_{\rm min}}.
\label{eq:pf}
\end{equation}
\noindent In the geometry that we are considering, the maximum and minimum
fluxes, $F_{\rm max}$ and $F_{\rm min}$, correspond to phases $\gamma=0$
and $\gamma=\pi$, respectively. Both fluxes are integrated over the given
energy bands.

The results of these fits show that it is not possible to reproduce
the observed modulation if the emission pattern of both the hot and
warm component is isotropic. The introduction of General Relativity
effectively suppresses the modulation to below that observed, for any
reasonable assumed NS radius. We ascribe the observed modulation to
additional anisotropic emission from the thermal regions and test this
assumption using a simple model of cosine beaming described in \S2.
For each value of the radius, we ran through the ($\psi$,$\xi$) grid
of models for different combinations of the beaming parameters $n_w$
and $n_h$ of the warm and hot components, respectively. More
specifically, we varied $n_w,n_h$ between 0 and 2, in increments of
0.5 (note that the softer component is less modulated than the harder
one). The fact that the warm component dominates in flux in the
softest energy band, while the hot component dominates in the hardest
energy band, allows us to constrain the degree of anisotropy of these
two components independently. For each set of angles $\psi$ and $\xi$
on the grid, we computed the reduced $\chi^2_\nu$ and kept track of its minimum.
Table~2 reports the minimum $\chi^2_\nu$ that was obtained for a few
representative values of the beaming parameters.
                                                   
With the introduction of beaming, we are able to identify a set of
model parameters that is able to reproduce the observed modulation in
the first 4 epochs. The main results from our modeling can be
summarized as follows:

\begin{itemize}
\item[{\em i)}]
The modulation of the hot spectral component requires an anisotropic
radiation pattern. For a { $\cos^{n_h}\delta$} emission profile, the best
match to the data is obtained with $n_h\approx$~1.
\item[{\em ii)}]
No similar beaming is required to model the warm component modulation
(i.e. $n_w\sim 0$).
\item[{\em iii)}]
We constrain the emission geometry by identifying allowed and forbidden
regions in the $\psi$-$\xi$ parameter space.
\item[{\em iv)}]
No NS radius is strongly preferred by the data. However, the range
of most preferred values of $\psi$ and $\xi$ varies with the radius
of the star.
\end{itemize}

\begin{figure}
\plotone{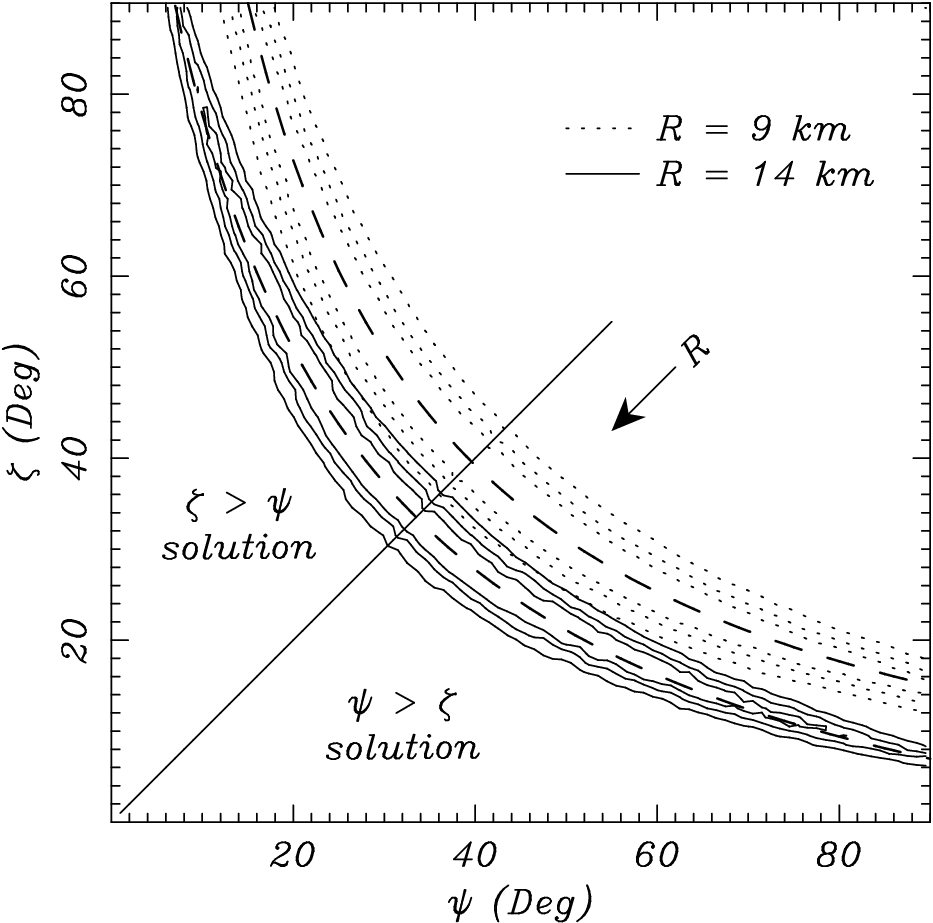}
\caption{Reduced chi-square ($\chi_{\nu}^2$) maps obtained by
comparing modulation data and model described in the text for a range
of viewing angles $\psi$ and $\xi$. The 68\%, 90\% and 99\% confidence
levels are shown for the best match to the observed pulsed fractions
using the beaming patterns $n_w = 0$, $n_h=1$, for $R=9$~km and $R=14$~km. 
The results are clearly degenerate with respect to an interchange of
$(\xi,\psi)$. For the adopted model, the locus of minimum $\chi_{\nu}^2$
depends only on radius and lies along a line
$(\xi+\beta(R))(\psi+\beta(R))\approx{\rm constant}$ ({\em dashed lines}). The
minimum $\chi^2_\nu$ for each value of the radius is reported in
Table~2, and the angles ($\psi,\xi$) to which this minimum corresponds
vary slightly with radius.}
\vspace{0.1in}
\label{fig:contour}
\end{figure}

For the optimal beaming parameters, $n_w=0$ and $n_h=1$, we plot the
$\chi^2_{\nu}$ map computed for the analysis of Table~2.
Figure~{\ref{fig:contour}} displays the 68\%, 90\% and 99\% confidence
levels drawn for the two extreme values of radii considered
($R=9,14$~km). As described above, this map is produced by comparing
the model and observed modulation over a range of possible
($\xi,\psi$) angle pairs, for our best fit spectral model
parameters. The range of allowed solutions defines a locus of points
in the $\xi-\psi$-plane along the line $(\xi+\beta(R))(\psi+\beta(R))=
c$, were $c$ is, {to zeroth order} a constant for the given model
{(it varies by $\la 9\%$ between the two extreme values of radii
considered here)}, while $\beta(R)$ contains the strong dependence on
the NS radius.  { A linear regression fit yields the relations
$\beta=1.64\;R_{\rm km}-6.16$ and $c=40.65\;R_{\rm m} +1935$, where
$R_{\rm km}$ is the radius of the star in km. }  The elongated shape of
the contour plots shows that the two angles $\psi$ and $\xi$ are
highly correlated in the fit. This is a result of the fact that the PF
depends on {a combination} of these two angles.

The dependence on the NS radius is clearly seen in
Figure~{\ref{fig:contour}}. For each given value of $\psi$, the best
fit loci moves toward larger values of $\xi$ as the radius of the star gets
smaller. The contour levels for intervening values of the radii fall
in-between those shown.  This trend with radius can be understood as
follows. As the radius decreases, the larger angular sizes and
temperatures conspire to decrease the PF for the same viewing angles
$\psi$ and $\xi$.  In addition, the gravitational effect of light
bending reduces the modulation even further for small stars.  In order
to reproduce the same observed PF for a given $\psi$, 
a correspondingly larger $\xi$ is therefore needed for
smaller stars since, for any given value of each of these two angles,
a larger value of the other produces a larger modulation. This is the
reason for the shift of the confidence levels toward larger values of
$\xi$ for a given $\psi$, when the radius of the star gets smaller.

\begin{figure*}
\includegraphics[angle=270.0,width=1.0\linewidth,clip=]{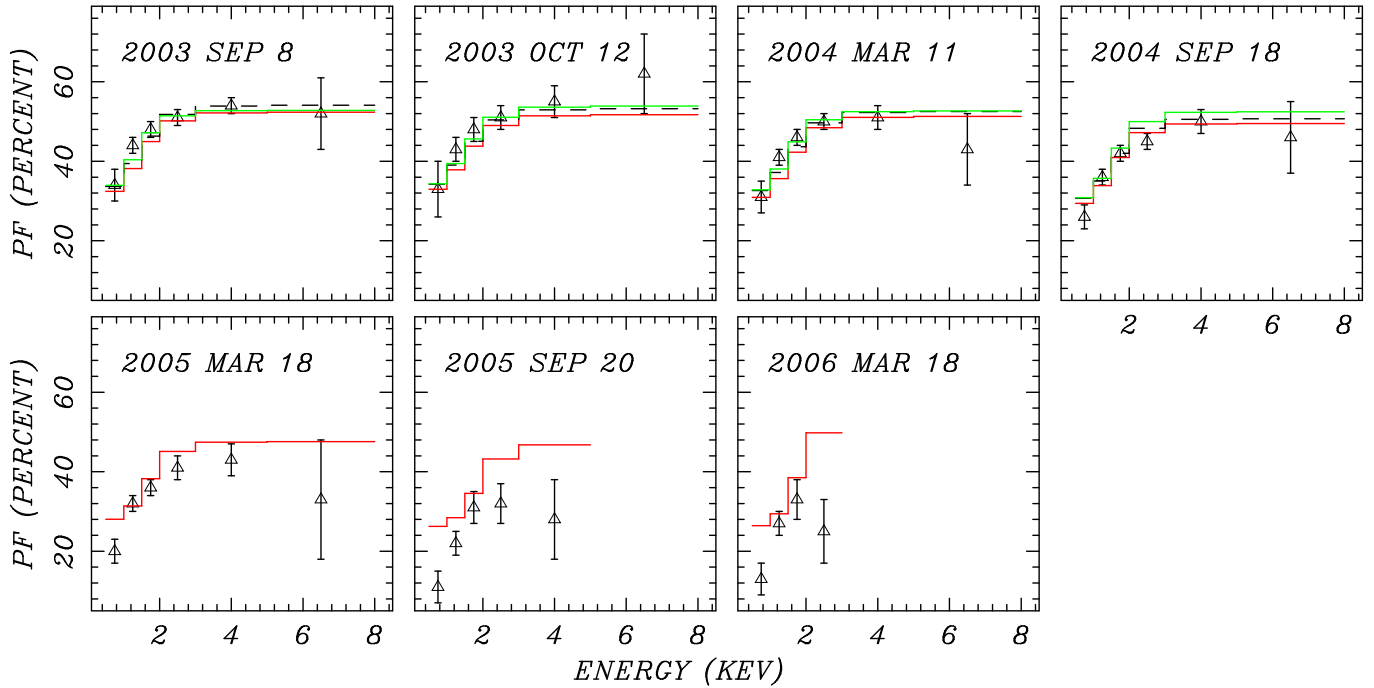}
\caption{Comparison between the measured pulsed fraction (PF) and that
predicted by our model for one particular set of best-fit parameters
determined for a $R=12$~km neutron star. Results are shown for the
best fit model using all seven data sets (red line; $\psi=49^{\circ}$
and $\xi=24^{\circ}$) and for the first four data sets only (green
line; $\psi=53^{\circ}$ and $\xi=23^{\circ}$).  At the later times,
the model is seen to deviate significantly from the data.  A likely
explanation is the increased contribution of the unmodeled emission
from the stellar surface over time compared to the modeled flux.  In
the first 4 data sets, the {\it dashed line} is to be compared to the
{\it green line}, showing the effect of using the face-on spectra
instead of the iterated spectra for the given viewing geometry (see
text for details). {Each data point is drawn in the middle of the corresponding
energy band, for those bins with sufficient photons. } }
\vspace{0.1in}
\label{fig:pf}
\end{figure*}

Figure~{\ref{fig:pf}} compares the data and model modulation for the
($\psi,\xi$) values that yield minimum $\chi^2_\nu$ for the case
$R=12$ km, $n_w=0$, $n_h=1$. We show the results for two cases, one
using all seven epoch data sets, and one for the case of the
first four data sets only. In the former case, the minimum corresponds
to $\psi=49^\circ$ and $\xi=24^\circ$, while in the latter, it occurs for
$\psi=53^\circ$ and $\xi=23^\circ$.  We find excellent agreement between data
and model modulation for the first four observations alone; however,
the later data sets show increasing discrepancies, noticeably
increasing the overall contribution to $\chi^2$.  This confirms the
suspicion that, at later times, the emitted radiation acquires an
unpulsed (or very mildly pulsed) contribution from the surface of the
star.  Our spectral fits, as described above, do not resolve this
underlying stellar component, and therefore the spectral parameters
close to quiescence might not be as representative of the underlying
physical parameters of the system. However, the constraints on the
viewing and emission geometry that we derive using the first 4 data sets
alone (cfr. Table 2 and Fig.2) can be considered robust, since the early
data sets are basically unaffected by the presence of the star underlying
emission.

\begin{figure*}[t]
{\includegraphics[angle=270,width=\textwidth]{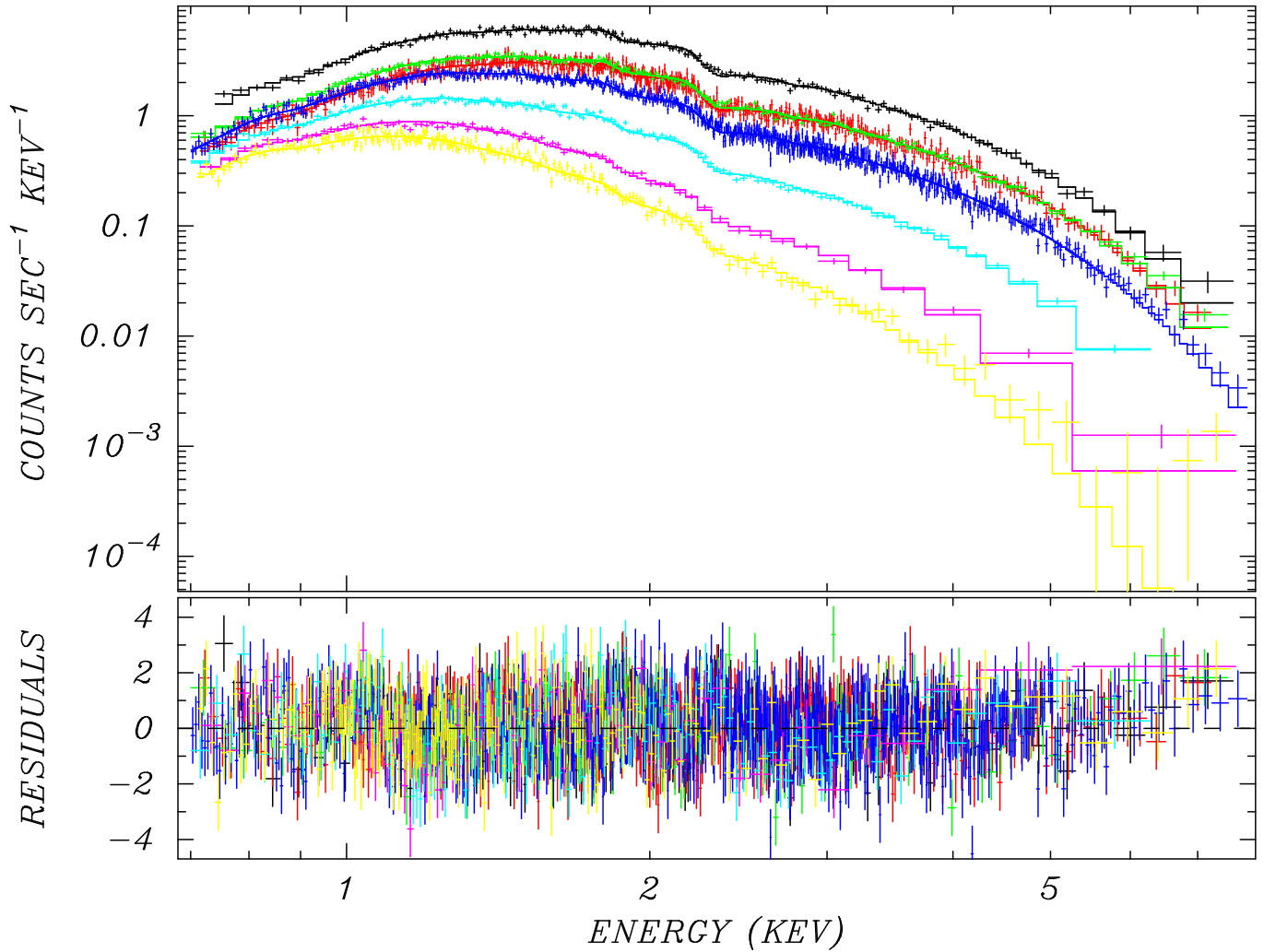}}
\caption{\xmm\ {phase-averaged} spectra of \taxp\ obtained at 7 epochs fitted with the
model presented in the text, for the specific case of $R=12$ km,
$\psi=53^\circ$ and $\xi=23^\circ$. The parameters for the best fit model
are reported in Table~3.  The lower panel shows the collected
residuals to this fit for each spectrum.}
\label{fig:spectrum}
\end{figure*}

As a final step in our analysis we consider the validity of our
initial method of assuming a face-on spectrum to derive the spectral
model parameters that are then used to compute the
modulation. As discussed in \S3.1, the original fits
were generated assuming $\psi=\xi=0$ for simplicity, prior to
determining the observational geometry of the pulsar system.  We now
show that this is an excellent assumption by refitting to the data the spectrum
assuming the specific case of $\psi=53^{\circ}$ and $\xi=23^{\circ}$,
and then recomputing the modulations. The best fit spectral
parameters are reported in Table~3 and shown in
Figure~{\ref{fig:spectrum}}. For these spectral parameters the model
modulation is shown in Figure~{\ref{fig:pf}} as the dashed-line for
the case of the first four data sets. The results are identical within
the statistical uncertainty in the data.

{Similarly, we performed a test in order to assess the validity of
our method of analysis which separated the spectral and timing studies
and used, as spectral parameters for the timing analysis, those
obtained from a phase-averaged fit to the spectrum. We again
considered the specific case of $\psi=53^{\circ}$ and
$\xi=23^{\circ}$, and extracted phase-resolved spectra by dividing the
observed spectrum into 5 equally spaced bins.  We then fitted the
model to the two bins centered on the maximum and minimum of the flux,
respectively. We found that the temperature at flux minimum is lower than
the temperature at flux maximum by $ \la 10\%$ for all XMM epochs
with the exception of the first one, where the difference is $\sim
20\%$. However, this difference is within one $\sigma$ of the combined
uncertainty in temperature for all epochs. Therefore, our adopted
method is quite robust within the statistical uncertainty of the
data.}

\begin{deluxetable*}{lccccccc}
\tablecolumns{8}
\tighten
\tablewidth{0.0pt}
\tablecaption{\bf  Spectral Results for Radius $\bf R=12$ km and Viewing Angles $\bf \psi=53$ deg, and $\bf \xi=23$ deg; $\bf D=3.3$~kpc, $\bf N_{\rm H}=6.8\times 10^{21}$ cm$^{-2}$\label{tab:specone}}
\tablehead{
\colhead{Parameter} &  \multicolumn{2}{c}{2003}  & \multicolumn{2}{c}{2004} & \multicolumn{2}{c}{2005} & \colhead{2006}\\ 
\colhead{}          &  \colhead{Sep 8} &  \colhead{Oct 12}  & \colhead{Mar 11} & \colhead{Sep 18} & \colhead{Mar 18} & \colhead{Sep 20} & \colhead{Mar 12} 
}
\startdata
$kT_h$ (keV)   & $0.84^{+0.03}_{-0.02} $ & $0.87^{+0.04}_{-0.04}$ & $0.85^{+0.01}_{-0.03}$& $0.82^{+0.02}_{-0.02}$& $0.73^{+0.03}_{-0.02}$& $0.63^{+0.03}_{-0.04}$& $0.53^{+0.04}_{-0.03}$\\
$kT_w$ (keV)  & $0.30^{+0.68}_{-0.03}$ & $0.34^{+0.11}_{-0.06}$  & $0.31^{+0.01}_{-0.03}$  &$0.29^{+0.04}_{-0.04}$& $0.26^{+0.06}_{-0.02}$& $0.23^{+0.01}_{-0.03}$& $0.21^{+0.02}_{-0.01}$\\
$\beta_h$ (deg)	 & $8.0 ^{+0.2}_{-0.2}$   & $7.1^{+0.7}_{-0.2}$   & $5.5^{+0.3}_{-0.1}$  & $4.52^{+0.06}_{-0.08}$  & $3.8^{+0.4}_{-0.2}$   & $3.02^{+0.04}_{-0.05}$   & $3.4^{+0.5}_{-0.8}$\\
$\beta_w$ (deg)   & $39^{+2}_{-2}$  & $30^{+2}_{-3}$   & $31.0^{+0.7}_{-1.7}$  & $33.9^{+0.5}_{-1.0}$  & $37.7^{+0.7}_{-0.1}$   & $44^{+1}_{-3}$   & $60^{+5}_{-3}$
\enddata
\tablecomments{Uncertainties in spectral parameters are 90\% confidence for two interesting parameters. The $\chi^2_\nu$ of the fit is 1.08 for 1692 dof.}
\vspace{0.2in}
\end{deluxetable*}

\section{Discussion}

The time-dependent spectrum and pulse modulation of the transient
magnetar \taxp\ provide a unique diagnostics of its emission
properties and geometry. Under the assumption that the post-burst
emission is described by two thermal components, as early analysis of
this object suggested, we have been able to extract information on
some of the physical properties of the star, through a detailed
modeling of the combined spectra and pulsed modulation together.

We found that, while the phase-averaged spectral fits alone are
degenerate with respect to the emission pattern of the radiation,
including modeling of the energy-dependent pulsed flux allows us to
constrain the properties of the emission region.  In particular, since the warm
component dominates in the lowest energy band, while the hot component
dominates in the highest energy band, the PFs are able to determine
the degree of anisotropy of these two components independently.  We
found that the warm component is best described by an isotropic
emission pattern, while the hot component is well represented by an
emission pattern of pencil type, {$f(\delta)\propto \cos\delta$,
where $\delta$ is the angle that the emitted photons make with respect
to the normal to the surface of the star}.  The different type of
radiation pattern required by the low and the high energy components
could be seen as an indirect confirmation of our assumption that the
contribution from these two energy bands does indeed come from
different components. Beaming of radiation in the direction of the
magnetic field is predicted by models of magnetized atmospheres in the
limit of high magnetic fields (e.g. van Adelsberg \& Lai 2006).  Since
the hot spot is produced in a region much smaller than that of the
warm component, it is more likely to find a configuration with
parallel field lines in the hot region (and most likely perpendicular
to the surface, which favors the heat flow), than in the warm
region. The latter might rather encompass regions with different
orientations of the magnetic field, hence resulting in an overall more
homogeneous radiation pattern.

The strongest constraints that we derived from our analysis are on the
geometry of the star. In particular, we determined the allowed regions
for the angle ($\xi$) between the spot axis and the rotation axis, and
the angle ($\psi$) between our line of sight  and the
rotation axis. These two angles determine the minimum and maximum
angles between the line of sight and the spot axis, given respectively
by $\alpha_{\rm min}=\xi-\psi$ and $\alpha_{\rm max}=\xi+\psi$.  We
find that, while the range of $\alpha_{\rm min}$ is compatible with
very small angles (including zero), however $\alpha_{\rm max}$ must
always be large, $\ga 60^\circ$ within 3$\sigma$ confidence level for
any value of the star radius.  Being able to rule out to a high
confidence level small viewing angles $\alpha$ for the entire rotation
period of the star bears important implications for models of the
observed radio emission from this object. In fact, Camilo et
al. (2007a) showed that the peaks of the radio and the X-ray pulses
are aligned, suggesting that the footpoints of the active magnetic
field lines on which radio emission is generated are also the
locations of the concentrated heating that is responsible for the
enhanced X-ray emission, {This means that, even if the radio
emission is likely produced at much higher altitudes on the surface of
the star than the X-ray emission, however the axis where the two
emissions peak is the same (or very close).}

Attempts to constrain the viewing angles of \taxp\ using radio
polarimetry were made by Camilo et al. (2007b). They found two
configurations likely, one with $\xi\sim 70^\circ$ and $\alpha_{\rm
min}\sim 20^\circ-25^\circ$, and another with $\xi\sim 4^\circ$ and
$\alpha_{\rm min}\sim 4^\circ$. Our fits rule out the second
configuration to a high significance level.  The value of $\alpha_{\rm
min}\sim 20^\circ-25^\circ$ on the other hand is perfectly compatible
with our results, albeit it requires $\xi\sim 60^\circ$ if $R=9$~km
and $\xi\sim 50^\circ$ if $R=14$~km. Although our confidence levels
are close to one of the two solutions of Camilo et al. (2007b), we
cannot make a formal statistical comparison with their results, since
they do not have a reliable estimate of the parameter uncertainties
from their radio measurements (F. Camilo, priv. comm.).

As discussed by Camilo et al. (2007b), the observed wide radio pulse
profile of $\approx 0.15\;P$ can be explained by either a model in
which the magnetic and rotation axes are almost aligned, or by a model
in which the emission height is very large.  Our results strongly rule
out the first scenario, hence implying a large emission height. This,
in turn, implies a large opening angle of the beam (Gil et al. 1984),
comparable to that observed in young pulsars (Johnston \& Weisberg
2006). These characteristics of the radio emission, if common in
magnetars, make more stringent the limits on the radio for the
greatest majority of the objects that have not been detected in this
waveband, and leave even more open the question of what is that makes
some magnetars different.

\acknowledgements 
We thank Jules Halpern, Andrei Beloborodov and
Fernando Camilo for stimulating discussions on several aspects of this
work.  RP thanks Columbia University for the kind hospitality during
the several visits made while this work was carried out. 
{We also thank the referee for his/her insightful and helpful comments
on our manuscript.}

\end{document}